# Observations and Analysis of Absorption Lines including *J* = *K* Rotational Levels of $CH_3CN$: The Envelope of Sagittarius B2(M)

Mitsunori Araki,[1,2] Shuro Takano,[3] Nobuhiko Kuze,[4] Yoshiaki Minami,[1] Takahiro Oyama,[1] Kazuhisa Kamegai,[5] Yoshihiro Sumiyoshi,[6] and Koichi Tsukiyama[1,2]

[1] Department of Chemistry, Faculty of Science Division I, Tokyo University of Science, 1-3 Kagurazaka, Shinjuku-ku, Tokyo, 162-8601, Japan; araki@rs.tus.ac.jp
[2] Research Institute for Science and Technology, Tokyo University of Science, 2641, Yamazaki, Noda, Chiba, 278-8510, Japan
[3] Department of Physics, General Studies, College of Engineering, Nihon University, 1 Nakagawara, Tokusada, Tamuramachi, Koriyama, Fukushima, 963-8642, Japan
[4] Department of Materials and Life Sciences, Faculty of Science and Technology, Sophia University, 7-1 Kioi-cho, Chiyoda-ku, Tokyo, 102-8554, Japan
[5] Astronomy Data Center, National Astronomical Observatory of Japan, 2-21-1 Osawa, Mitaka, Tokyo, 181-8588, Japan
[6] Division of Pure and Applied Science, Graduate School of Science and Technology, Gunma University, 4-2 Aramaki, Maebashi, Gunma, 371-8510, Japan

**ABSTRACT**
Molecules in diffuse and translucent clouds experience cooling as a result of radiation and less excitation from collisions. However, a rotation around a molecular axis of acetonitrile, $CH_3CN$, cannot be cooled by radiation, causing rotational populations to concentrate at the $J = K$ levels. We aim to search for absorption lines of $CH_3CN$ having $J = K$ level concentrations in diffuse and translucent clouds. The $J_K = 4_3–3_3$ transition at 73.6 GHz was investigated toward Sgr B2(M) in the Galactic Center region and other sources, using the Nobeyama 45 m telescope. Based on the detected absorption lines toward Sgr B2(M), a radiation temperature of 2.8 ± 0.5 K, kinetic temperature of 88 ± 29 K, and column density of (1.35 ± 0.14) × $10^{14}$ $cm^{-2}$ were derived for this molecule, revealing extremely concentrated $J = K$ levels due to the lower excitation temperature and the higher kinetic temperature. The absorption lines occurred at a velocity of 64 km $s^{-1}$. The results confirm that $CH_3CN$ with $J = K$ level concentrations exists in the envelope of Sgr B2(M).

## 1. INTRODUCTION

The origin of complex organic molecules (COMs) is one of the most important questions in astrochemistry and astrophysics. Naturally enough, distributions of COMs in space provide clues for revealing their origin. There is no absolute definition of COM size and, in this study, we use the definition of ≥ 6 atoms (e.g., Herbst & Van Dishoeck 2009). In both the Galactic Center and the Galactic Disk, many COMs have been found in star-forming regions, such as Orion KL (e.g., White et al. 2003), Sagittarius (Sgr) B2 cores (e.g., Belloche et al. 2013), and IRAS 16293-2422 (e.g., Jørgensen et al. 2016), and in dense clouds without signs of star formation, such as the typical Galactic Center clouds reported by Requena-Torres et al. (2006, 2008).

Relatively small COMs, such as $CH_3OH$, $CH_3CN$, $CH_3CHO$, and $NH_2CHO$, have been detected in translucent clouds (Thiel et al. 2017, 2019), and CNCHO having four heavy atoms was discovered in the spatially extended region toward Sgr B2(N) (Remijan et al. 2008). Conversely, in diffuse clouds, the large molecules found, *cyclic*-$C_3H_2$ (Cox et al. 1988), $H_2CCC$ (Liszt et al. 2012), and $CH_3CN$ (Liszt et al. 2018), consist of five or six atoms, except for $C_{60}^+$ (Campbell et al. 2015, 2016, Walker et al. 2015). Liszt et al. (2018) reported that $CH_3CN$ is ubiquitous in the local diffuse molecular interstellar medium.

The question is whether COMs are abundant and difficult to detect in diffuse and translucent clouds, or they are not abundant. One potential candidate for investigating this question is the Sgr B2 region. The structural model of the Sgr B2 region reported by Hüttemeister et al. (1993, 1995) consists of three cores (N, M, and S) within a moderate-density region (hereafter the moderate-density envelope in this paper). The three cores, with densities $n(H_2) \sim 10^7$ $cm^{-3}$, are embedded in the moderate-density envelope ($n(H_2) \sim 10^5$ $cm^{-3}$) with an extended lower density envelope ($n(H_2) \sim 10^3$ $cm^{-3}$) further from the cores.

Although further detection of COMs by radio observations in diffuse and translucent clouds is anticipated, the search for emission lines from COMs is difficult. This is attributed to molecules with permanent dipole moments being cooled by radiation and less frequent collisions in diffuse and translucent clouds, making emission lines difficult to detect. Therefore, the search for absorption lines is an alternative option for detecting COMs in diffuse and translucent clouds. Bright background radio continuum sources, such as the Sgr B2 region, are required to observe the absorption lines of molecules. This requirement further restricts the detection of COMs in diffuse and translucent clouds.

Several COMs detected toward Sgr B2(N) demonstrate absorption lines at a velocity of 64 km s$^{-1}$ (Hollis et al. 2004a, 2006a, 2006b, Loomis et al. 2013, Zaleski et al. 2013, McGuire et al. 2016), and $H_2COH^+$ (Ohishi et al. 1996) and other small molecules (e.g., Hüttemeister et al. 1995) indicate absorption lines near 64 km s$^{-1}$. These reports suggested that almost all COMs with absorption lines caused by continuum emission from the cores are located in the envelope, all having the same velocity of 64 km s$^{-1}$. In this study, we selected the same line of sight toward Sgr B2(M) as our main source, for two reasons: Emission lines of the Sgr B2(M) core have a velocity of 62 km s$^{-1}$, which facilitates distinguishing absorption lines from emission lines, and Sgr B2(M) is optically thin compared with Sgr B2(N) (Belloche et al. 2013).

Absorption lines for the larger COMs would be weak, because intensities of absorption lines are almost proportional to population differences between upper and lower states of transition. Conversely, the stronger absorption lines of small COMs facilitate their detection. The smallest COM has six atoms and is considered appropriate for tracing COMs. Some well-known COM-rich clouds, such as Orion KL (e.g., White et al. 2003), IRAS 16293-2422 (e.g., Cazaux et al. 2003), and the Sgr B2 cores (e.g., Belloche et al. 2013), are rich in acetonitrile ($CH_3CN$), and this molecule has a large permanent dipole moment of 3.9220 D (Gadhi et al. 1995). Hence, $CH_3CN$ can be a possible indicator of COMs in diffuse and translucent clouds.

Molecules in interstellar space are cooled by radiation and excited by collisions with hydrogen molecules. Generally, the excitation temperatures of molecules with dipole moments result from the competition between cooling and excitation, whereas the excitation temperatures of molecules without dipole moments approximately match the kinetic temperature of hydrogen molecules in a cloud, because of the lack of radiational cooling. However, for molecules with a three-fold axis of symmetry, such as $NH_3$, $H_3O^+$, $CH_3CN$, and $CH_3CCH$, overall rotation ($J$ rotation, where $J$ is a quantum number of a total angular momentum) is cooled by the radiation of $J + 1 \rightarrow J$ transitions. The rotation around its molecular axis ($K$ rotation, where $K$ is the projection of a total angular momentum along the molecular axis, $K \leq J$) is maintained because only $\Delta K = 0$ transitions are allowed in the emission and absorption of radio waves. Therefore, an excitation temperature of the $K$ rotation is higher than that of the $J$ rotation, i.e., populations are concentrated at the $J = K$ rotational levels. The molecular axis can be regarded as a *hot axis*. In this paper, we have named this concentration at the $J = K$ rotational levels in the population of a prolate molecule the *hot axis effect* for convenience, where these levels correspond to the so-called *metastable levels* in the case of oblate small molecules such as $NH_3$ (Walmsley & Ungerechts 1983) and $H_3O^+$ (Lis et al. 2014). In diffuse and translucent clouds, this sub-thermal condition is obtained under a condition of less collisional excitation because of their low-density conditions. The excitation temperature of the $K$ rotation is approximately equal to the kinetic temperature of the cloud, whereas the excitation temperature of the $J$ rotation is close to the radiation temperature of the cloud. The concentrations of the populations in the $J = K$ levels induce clear absorption lines of $J + 1 \leftarrow J$ and $\Delta K = 0$ transitions. This effect is an essential factor in evaluating accurate abundances of $CH_3CN$ in diffuse and translucent clouds.

Candidates of such absorption lines have been reported. A spectral dip at the frequency of the $J_K = 4_3$–$3_3$ transition for $CH_3CN$ was observed toward Sgr B2 by Cummins et al. (1983), who suggested that the dip appears to be an absorption and may be produced by the lower-excitation foreground material. Typical absorption lines of $CH_3CN$ were detected in the $J_K = 6_5$–$5_5$ and $5_4$–$4_4$ transitions toward Sgr B2(M) by De Vicente et al. (1997) and Belloche et al. (2013). These lines were explained by a hot diffuse envelope having a kinetic temperature of 300 K. If these absorption lines are the result of the $J = K$ level concentrations by the hot axis effect, a $J$-rotation temperature could be close to the cosmic background temperature. However, in this envelope, $J$-rotation temperature could not be determined. This is because the line profiles were very complex and the only detected absorption lines were from $J = K$ levels (De Vicente et al. 1997). To find the $J = K$ level concentrations by the hot axis effect, absorption lines from all the $K$ levels need to be evaluated.

The conditions of the emission component toward Sgr B2 have been derived as 85 ± 10 and 16–17 K for kinetic and $J$-rotation temperatures, respectively (Cummins et al. 1983), and similar conditions have been found in the Orion Molecular Cloud (Hollis 1982, Andersson et al. 1984, Loren & Mundy 1984). However, the $J$-rotation temperatures reported in these papers are still higher than the cosmic background temperature. When radiational cooling is more effective, $J$-rotation is further cooled. As a result, $J = K$ level concentrations are expected to clearly appear, i.e., the $J$-rotational temperature is measured to be approximately 2.7 K. Definite detections of these concentrations have yet to be reported.

This study aims to identify $J = K$ level concentrations of $CH_3CN$ by the hot axis effect through observations and analysis of absorption lines from all the $K$ levels in the Sgr B2 region. In Section 2, observations of $CH_3CN$ and $CH_3CCH$ using the Nobeyama 45 m telescope are described, where $CH_3CCH$ is included to provide a kinetic temperature that can be used to analyze rotational profiles of $CH_3CN$. In Section 3, we report the detection of absorption lines of $CH_3CN$ in the envelope toward Sgr B2(M).

## 2. OBSERVATIONS

The Nobeyama Radio Observatory 45 m telescope[1] was employed in this work. To detect $CH_3CN$ in diffuse and translucent clouds, we searched for absorption lines mainly toward Sgr B2(M), but we also observed the quasar

[1]The 45 m telescope is operated by the Nobeyama Radio Observatory, a branch of the National Astronomical Observatory of Japan.

**Table 1**
Observation Positions

| | RA (J2000) | Dec. (J2000) | Off position [a] RA | Off position [a] Dec. | Pointing | Date |
|---|---|---|---|---|---|---|
| B0212+735 [b] | $2^h\ 17^m\ 30^s.8$ | 73° 49′ 32.6″ | +10′23″ | −1′33″ | TX Cam | Mar and Apr 2016 |
| Orion IRc2 | $5^h\ 35^m\ 14^s.5$ | −5° 22′ 29.6″ | +15′ | +20′ | Orion KL | Mar and Apr 2016 |
| Sgr B2(M) | $17^h\ 47^m\ 20^s.4$ | −28° 23′ 07.3″ | +60′ | 0′ | VX Sgr | Jan 2018 [c] |
| Sgr B2(M) edge [d] | $17^h\ 47^m\ 24^s.0$ | −28° 22′ 15.4″ | +60′ | 0′ | VX Sgr | Feb 2018 |
| W49N [e] | $19^h\ 10^m\ 13^s.2$ | 9° 06′ 12.0″ | +14′ | +14′ | GL 2445 | Feb 2017 [f] |
| W51 [e] | $19^h\ 23^m\ 43^s.9$ | +14° 30′ 36.4″ | +30′ | 0′ | RR Aql | Feb 2017 |

[a] for position switching.
[b] Observation position is from Liszt and Pety (2012). The position without CO emission in the map was selected as the off position.
[c] Feb 2018 for $HC_3N$.
[d] Edge of Sgr B2(M), i.e., offset position from the center of Sgr B2(M).
[e] Observation position is from Sonnentrucker et al. (2010).
[f] including the short integrations in Jan 2018.

B0212+735 and the massive star-forming regions Orion IRc2, W49N, and W51. The position switching mode was selected for the observations. The observed clouds and their off positions are listed in Table 1. We searched in these clouds for the $J$ = 4–3 transitions of $CH_3CN$ at 73.6 GHz according to the presumption of intensities for absorption lines in Appendix 1. Based on reported observations (Cummins et al. 1983, De Vicente et al. 1997, Belloche et al. 2013), Sgr B2(M) was expected to be the best source for detecting absorptions. In the case of Sgr B2(M), the $J$ = 6–5 transition of $CH_3CN$ at 110.3 GHz was investigated towards the edge (+48″, −52″; 2.7 pc) of the Sgr B2(M) core (see Table 1), to detect emission lines that do not overlap with absorption lines; this observational point has little disturbance from absorption because of the small amount of background continuum emission. Assuming a widely uniform kinetic temperature such as with $NH_3$ (Hüttemeister et al. 1993), observations made toward the edge will have the same kinetic temperature as an emission component toward the Sgr B2(M) core, facilitating the extraction of absorption lines from superimposed spectra.

To evaluate the kinetic temperature of an emission component directly, we observed the $J$ = 5–4 transitions of $CH_3CCH$ at 85.4 GHz toward Sgr B2(M) (see Section 3.1), as the emission lines of these transitions would not be disturbed by the absorption lines of the envelope (Belloche et al. 2013). For comparison, we also observed the same transitions in B0212+735, Orion IRc2, W49N, and W51.

For observations in the 73.6 and 85.4 GHz frequency regions, T70H/V dual-polarization side-band separating SIS mixer receivers were used, and the spectra obtained from different polarizations were averaged. The main beam efficiency was 0.55 in the frequency regions of both receivers. (Beam sizes are listed in Table 2.) A SAM45 highly flexible FX-type spectrometer (Kuno et al. 2011, Kamazaki et al. 2012) was employed for the backend. A channel separation of 61.04 kHz was selected with the SAM45, giving a bandwidth of 250 MHz for each array. This channel separation corresponds to a velocity resolution of 0.50 km $s^{-1}$ at 73.6 GHz.

For observations in the 110.3 GHz frequency region, FOREST dual-polarization side-band separating SIS mixer receivers (Minamidani et al. 2016) were utilized, where the main beam efficiency was 0.43, and the beam size was 14.6″, the average of the H and V receivers. A channel separation of 61.04 kHz was selected with the SAM45, giving a bandwidth of 250 MHz for each array, corresponding to a velocity resolution of 0.33 km $s^{-1}$ at 110.3 GHz.

To obtain the normalized intensities of absorption spectra, the continuum intensities, $T_C$, toward the target sources were observed at 80 GHz using the FOREST receivers, because T70H/V receivers do not support continuum observations. The observed intensities were calibrated using the standard source NGC7027 with intensities of 5.5, 5.0, and 4.6 Jy at 23, 43, and 86 GHz, respectively (Tsuboi et al. 2008). NGC7027 was observed on the same day as the observations of the 73.6 and 85.4 GHz frequency regions. For the 70–110 GHz region, it is expected that the continuum emission will be dominated by free-free emission from the HII regions and will not vary much with the frequency. Although influence of dust thermal emission, which varies with the frequency, increases in a higher frequency region, this can be still small in this frequency region. Thus, we use a constant value for the flux density. Because the source size of continuum is smaller than the beam size of the present observations (Akabane et al. 1988, Liu & Snyder 1999), $T_C$ can be constant. Hence, $T_C$ from 73.6 to 110.3 GHz of Sgr B2(M) is estimated from the observed flux density at 80.0 GHz, as listed in Table 2.

Telescope pointing was verified by observing the nearby SiO maser ($v$ = 1, $J$ = 1–0) sources every 30–90 min. A typical pointing deviation was a few arcseconds. The intensity scale was calibrated using the chopper wheel method.

## 3. RESULTS AND DISCUSSION

The emission lines of $CH_3CN$ and $CH_3CCH$ were observed toward the massive star-forming regions Orion IRc2, Sgr B2(M), W49N, and W51. For $CH_3CN$, a superimposed spectrum of the absorption lines from the envelope and the emission lines from the core was observed toward Sgr B2(M). Extracting the absorption lines requires an estimated profile of the emission lines from the core; this, in turn, requires the kinetic temperature of the core, which can be obtained from the profile of the emission lines of $CH_3CCH$, as described in Section 3.1. Extracted pure absorption profiles of $CH_3CN$, described in Section 3.2, are used for obtaining the kinetic temperature and radiation temperature in the envelope, as well as the column density of $CH_3CN$, as discussed in Section 3.3.

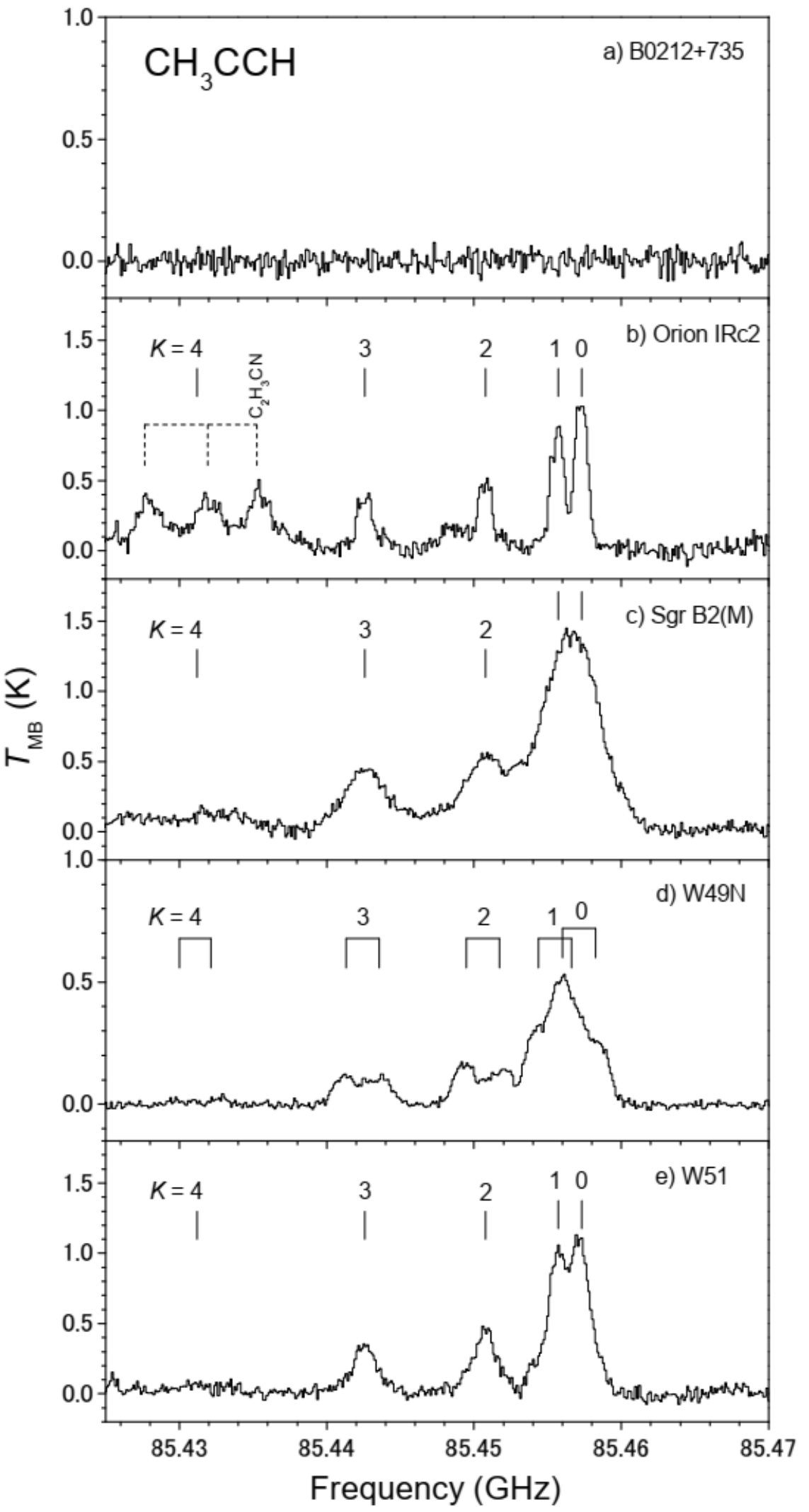

**Figure 1.** Emission lines of the $J_K = 5_K$–$4_K$ rotational transitions of $CH_3CCH$. The velocities of 3.5, 8.7, 62.0, 7.0, and 56.4 km $s^{-1}$ are assumed in the spectra of B0212+735, Orion IRc2, Sgr B2(M), W49N, and W51, respectively.

$HC_3N$ absorption lines are analyzed in Appendix 2: They were also observed, as other molecules with absorption facilitate the study of COM abundances. The abundances of $CH_3CN$, $HC_3N$, and other COMs in the various clouds are compared and discussed in Appendix 3.

### *3.1. Emission lines of $CH_3CCH$*

The profile of the emission lines of $CH_3CN$ must be analyzed using the kinetic temperature, excitation temperature of the $J$ rotation, and column density. In this study, the kinetic temperature was evaluated by a $K$-structure profile of the emission lines for $CH_3CCH$, assuming that the foreground $CH_3CCH$ does not produce absorption lines and that both molecules exist in the same region in the dense cloud. The emission lines in the $J$ = 5–4 transition of $CH_3CCH$ toward Sgr B2(M) are suitable for this evaluation, because the absorption lines of the envelope are not overlapped on these lines (Belloche et al. 2013). In addition, the velocity of $CH_3CCH$, about 62 km $s^{-1}$, is equal to that of the typical emission lines of the Sgr B2(M) core (Belloche et al. 2003).

Prior to the analysis of the $CH_3CN$ spectra, we analyzed the spectra of $CH_3CCH$ as illustrated in Figure 1. Toward Orion IRc2, Sgr B2(M), W49N, and W51, emission lines were detected, and absorption lines were not; neither emission nor absorption lines were detected toward B0212+735. The parameters of the $CH_3CCH$ emission lines are listed in Table 3. A column density of each $K$-ladder is determined conventionally (Snyder et al. 2006) by assuming the excitation temperatures, $T_{ex}$, listed in Table 4. $T_{ex}$ is an excitation temperature of the $J$ rotation, as a result of radiational cooling and excitation from collisions. The total column density is obtained by summing the column densities of $K$ stacks for $K$ = 0–4 (Table 4). Kinetic temperatures are derived from the column densities of $K$ stacks, assuming populations are doubled in the $K$ = 3, 6, … levels. As an example, the kinetic temperature in the Sgr B2(M) core is derived as 59 ± 7 K from comparison of the integrated intensities of the $K$ structure, $W$ in Table 3. This temperature agrees both with the 70 K for $CH_3CCH$ and the 60 K for $CH_3CN$ reported by Belloche et al. (2013) and was, therefore, used for the kinetic temperature of the $CH_3CN$ emission component, as described in the next section.

### *3.2. Extraction of absorption lines of $CH_3CN$ toward Sgr B2(M)*

The observed spectra of $CH_3CN$ are illustrated in Figures 2 and 3. Only emission lines were observed in the spectra toward Orion IRc2, W49N, W51, and the edge of the Sgr B2(M) core. Neither emission nor absorption lines were detected toward B0212+735. The parameters of the emission lines of $CH_3CN$ are presented in Table 5. For Orion IRc2 and W49N, the velocities measured are not the same with those of $CH_3CCH$. $CH_3CN$ and $CH_3CCH$ would reside in different components in these clouds. Using the intensities of the emission lines, the kinetic temperatures and column densities were determined by assuming the excitation temperatures, $T_{ex}$, as listed in Table 4, and the doubled populations of the $K$ = 3, 6, … levels.

The spectra of $CH_3CN$ toward Sgr B2(M) indicate a superimposed structure of emission and absorption lines, which must be separated. For emission lines from the Sgr B2(M) core, we assume that the kinetic temperature of

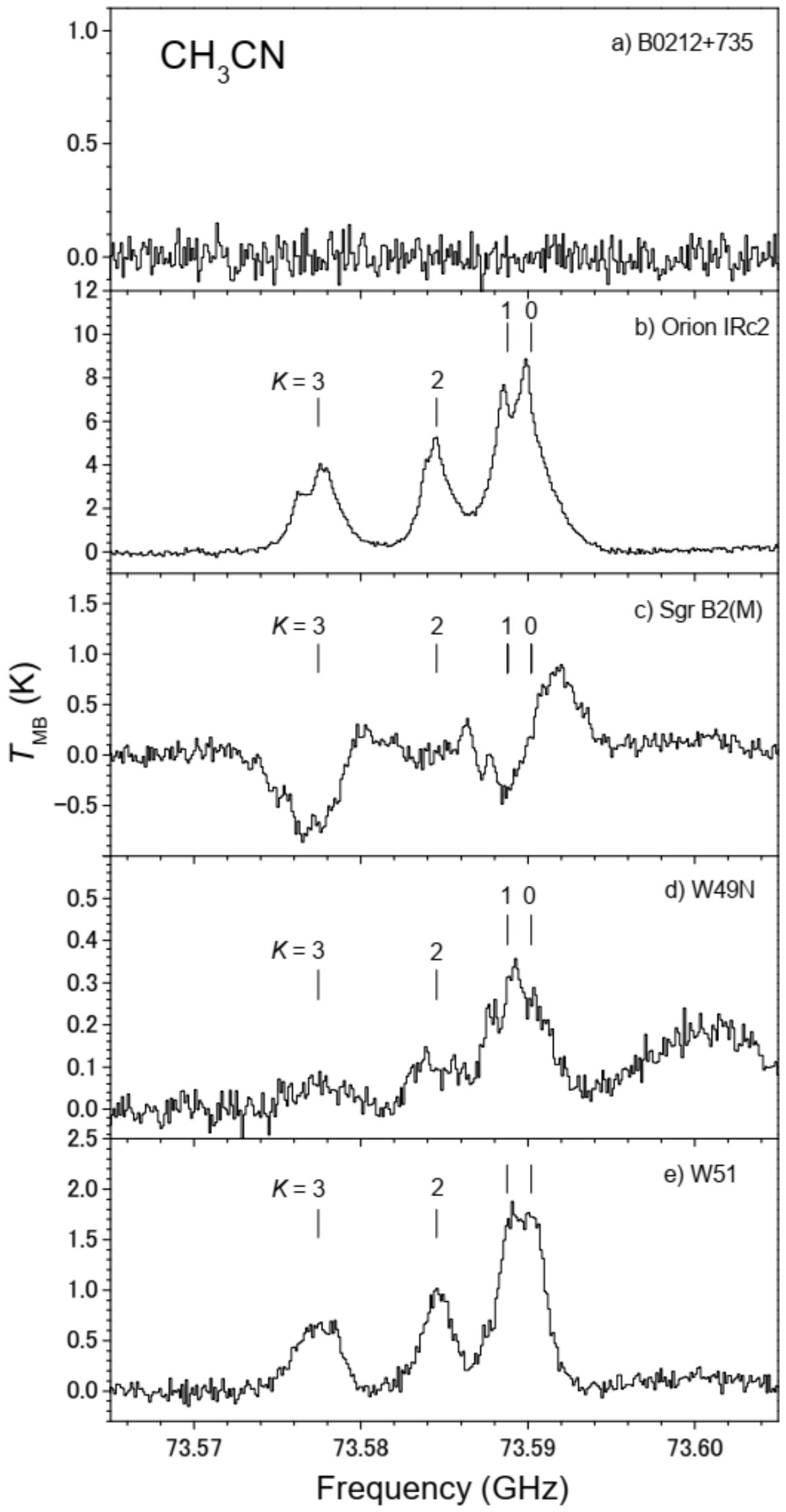

**Figure 2.** Lines of the $J_K$ = $4_K$ $-3_K$ rotational transitions of $CH_3CN$. The velocities of 3.5, 6.6, 62.0, 8.0, and 56.6 are assumed in the spectra of B0212+735, Orion IRc2, Sgr B2(M), W49N, and W51, respectively. A recombination line H63γ exists at 73.60 GHz (Lilley & Palmer 1968).

$CH_3CN$ is equal to that of $CH_3CCH$ (59 K, line 2 in Table 4) and that the excitation temperature of the $J$ rotation of $CH_3CN$ is 16 K (Cummins et al. 1983). The intensities of the emissions are determined to fit the line profile of the low-velocity side of the $J_K = 4_0$–$3_0$ transition in Figure 4(a); this is because the profile of the line is largely undisturbed by absorption lines, as the velocity and width of the absorption lines (64 and 14–15 km s$^{-1}$) are different from those of the emission lines (62 km s$^{-1}$ (Belloche et al. 2013) and approximately 20 km s$^{-1}$). Figure 4(a) also illustrates estimated emission profiles, where the column density is $1.7 \times 10^{14}$ cm$^{-2}$ in the Sgr B2(M) core. The same analyses can be applied to the reported spectra of the $J$ = 5–4 and 6–5 transitions. Using the same parameters as the case of $J$ = 4–3, a line profile of the low-velocity side of the $J_K = 5_0$–$4_0$ transition closely agrees with that reported by Belloche et al. (2013), as illustrated in Figure 4(b). This is the first validation for the estimated emission line profiles. The profile of $J$ = 6–5 (Belloche et al. 2013) is also reproduced, as illustrated in Figure 4(c), assuming a column density 1.35 times larger and the same temperature parameters. The requirement to use a larger column density may be because the beam observing the $J$ = 6−5 transition covers only the higher density region, owing to its smaller cross-sectional area.

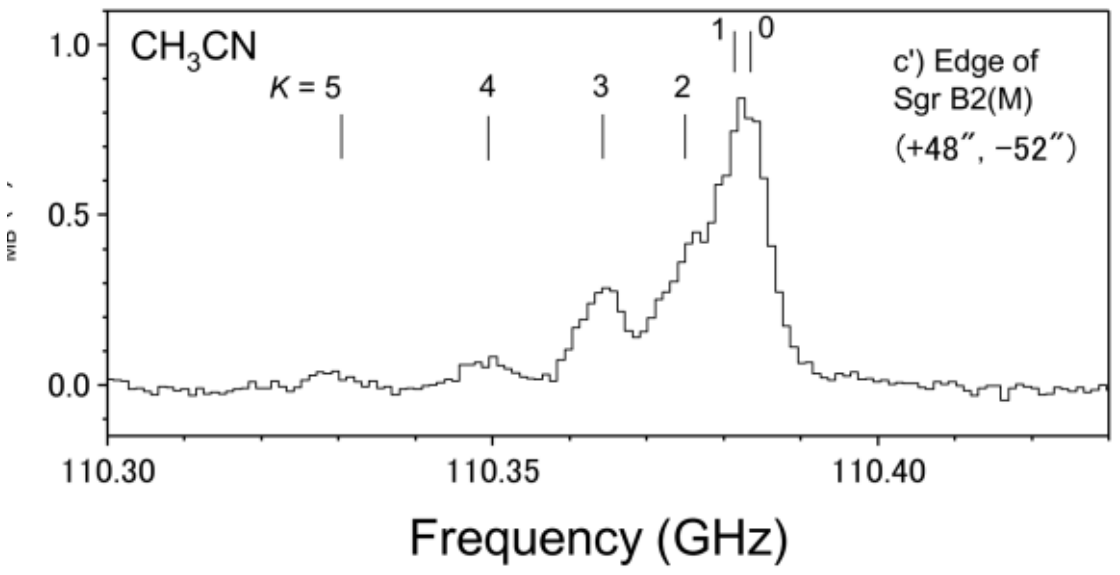

**Figure 3.** Emission lines of the $J_K$ = $6_K$–$5_K$ rotational transitions of $CH_3CN$ toward the edge of Sgr B2(M). A velocity of 62.0 km s$^{-1}$ is assumed.

To evaluate the validity of the above assumption, that the kinetic temperatures of $CH_3CN$ and $CH_3CCH$ in the Sgr B2(M) core are equal, we investigated the kinetic temperature through the $J$ = 6–5 emission line profile of $CH_3CN$ at the edge of Sgr B2(M). Fortunately, this profile indicates no absorption (Figure 3). The kinetic temperature is derived as 72 ± 18 K (line 8 in Table 4), which is consistent with the assumption that the kinetic temperatures at the edge and the core of Sgr B2(M) are equal.

Using the estimated emission profiles and interpolating the continuum intensities (Table 2), in Figure 4, we plotted absorption profiles with intensities, $1 + T_L/T_C$. In this process, $T_L$ (< 0) is the absorption intensity of a line, and $T_C$ in the present analysis is the sum of the line emission intensity and the continuum intensity, because the emission lines also contribute to background radiation. The parameters of the absorption lines are listed in Table 6. The absorption lines of $K$ = 2 and 3 for the $J$ = 5–4 transition overlap with the $HC_3N$ line for $v_5 = 1/v_7 = 3$ and an unidentified line around 91,980 MHz, respectively. Thus, the absorption lines of $K$ = 0, 1, 2, and 3 for $J$ = 4–3 and $K$ = 0, 1, and 4 for $J$ = 5–4 were selected to derive column density, radiation temperature, and kinetic temperature using the equations in Appendix 1. This is because these lines, in which profiles are close to Gaussian functions, are thought to be no blending with other lines.

### *3.3. Analysis of the absorption lines of $CH_3CN$ toward Sgr B2(M)*

The absorption lines of $CH_3CN$ were analyzed based on the theory considering radiational cooling and excitation from collisions for rotational distributions, described in Appendix 1. This method was proposed by Oka et al. (2013) and was modified to analyze non-linear molecules by Araki et al. (2014). First, it is necessary to estimate a rate for the collision-induced rotational transition, $C$, between $CH_3CN$ and $H_2$ using an initial presumption of the kinetic temperature $T_k$. Collisions of He, H, and electron with $CH_3CN$ were not considered because contribution of $H_2$ is dominant in a molecular cloud region. The value of $C$ is estimated using a hard-sphere collisional model, depending on $H_2$ densities and the $T_k$ of the cloud, as detailed in Table 7, where no $J$ and $K$ dependence is assumed. Since the dependence of $C$ on $T_k$ is weak compared to that on $H_2$ densities, three presumed values for kinetic temp erature, a) 10 K, b) 100 K, and c) 300 K, can support all cases of regions that include $CH_3CN$. Additionally, an approximate estimation of $T_k$ is sufficient to estimate $C$. An approximate value of $T_k$ can be estimated by comparing the intensities of an absorption line from a $J = K$ level and a blended absorption peak of $K = 0$ and 1. In case c), $T_k \geq 200$ K, the lines from $J = K$ (i.e., $J_K = 4_3$–$3_3$ and $J_K = 5_4$–$4_4$) are stronger than the blended peak of $K = 0$ and 1. This is not the case at present. Case a) would apply to starless dark clouds: 10 K is too low to be the temperature of the Sgr B2 region. Thus, in Table 7 we select case b), 100 K, as an approximate estimate of $T_k$ to derive a suitable $C$ value.

The radiation temperature, $T_r$, and $C$ are strongly correlated (see Equation (A1) in Appendix 1). In other words, as $C$ increases, it produces a similar effect as the increase of $T_r$ for rotational distributions, and either $T_r$ or $C$ can be determined from the relationship $T_L/T_C$ between the $J = 5$–4 and 4–3 transitions by fixing the other parameters. If $C = 1.4 \times 10^{-6}$ s$^{-1}$, $T_r$ is calculated as 2.7 K, which equals the cosmic background radiation. Thus, the value of $C$ must be $\leq 1.4 \times 10^{-6}$ s$^{-1}$ because $T_r \geq 2.7$ K. Where $T_k = 100$ K, $C$ has values of $1 \times 10^{-7}$ and $1 \times 10^{-6}$, with $H_2$ densities of $10^2$ and $10^3$ cm$^{-3}$, respectively (see Table 7). However, an $H_2$ density of $10^2$ cm$^{-3}$ is not suitable in the Sgr B2 region, since the densities of the envelopes were suggested to be $10^3$–$10^5$ cm$^{-3}$ by Hüttemeister et al. (1993, 1995). This suggests a value of $C = 1 \times 10^{-6}$ s$^{-1}$ for the present analysis.

The column density of $(1.35 \pm 0.14) \times 10^{14}$ cm$^{-2}$, radiation temperature of $T_r = 2.8 \pm 0.5$ K, and kinetic temperature of $T_k = 88 \pm 29$ K are derived as optimum values to reproduce the observed $T_L/T_C$ by least square fitting, using Equations (A1), (A2.1), (A2.2), and (A3), where errors come from the inconsistency in the intensities of the analyzed absorption lines. This inconsistency would be from imperfection of assumed cloud structure and the observational uncertainties of $T_L/T_C$ by weather condition, intensity calibration, and pointing.

An uncertainty of $T_C$ is thought to come from a bias by influence of dust thermal emission even though it is small. The observed flux density of Sgr B2(M) at 230.6 GHz was reported to be 44 Jy by Goldsmith et al. (1990). If $T_C$ suffers this bias, the continuum intensities from 73.6 to 110.3 GHz of Sgr B2(M) may be estimated from the observed values at 80.0 and 230.6 GHz assuming a linear relationship between $\log_{10}\nu$ and $\log_{10}F_\nu$, where $\nu$ is the frequency, and $F_\nu$ is the flux. In this case, $T_C$ increases by +0.1 K at 73.6 GHz ($J = 4$–3) and −0.1 K at 92.0 GHz ($J = 5$–4), and then $T_k$ and $T_r$ vary by 8 and 5%, respectively. On the other hand, the column density varies by −3% if $T_C$ in a whole frequency region uniformly changes by +0.1 K.

Influence of the value of $C$ on the results is also examined. Even though $C$ is assumed to be half of that used in the above analysis (i.e., $C_{half} = 0.5 \times 10^{-6}$ s$^{-1}$), the column density, $T_k$, and $T_r$ are merely increased to be 0.5, −3, and 3%, respectively. Thus, under the restriction of $C \leq 2 \times 10^{-6}$ s$^{-1}$ because of $T_r \geq 2.7$ K, an uncertainty of $C$ does not give large influence on the results.

At this temperature, an extremely low $T_{ex}$ is estimated, e.g., 3.0 K, which is between the $J_K = 4_3$ and $3_3$ levels. The extracted absorption profiles of $J = 4$–3 and 5–4 are reproduced by using the three derived parameters, depicted as blue lines in the lower panels in Figure 4. The kinetic temperature of $T_k = 88 \pm 29$ K is obtained from relative intensities of the $K$ structure and agrees with the initial estimate of $T_k = 100$ K used for estimating the $C$ value.

Line width of the emission component (approximately 20 km s$^{-1}$) is wider than that of the absorption component (12–15 km s$^{-1}$), and velocities of both the components are not the same (Tables 5 and 6). If $T_k$ of the emission component is over- or under-estimated, the extracted absorption curves of the $K = 2$ and 3 lines for the $J = 4 - 3$ transition cannot be reproduced. This is not the case at present (Figure 4). Following the first validation in Section 3.2, the absorption curves reproduced are the second validation of the estimated emission profiles. Thus, the assumption of $T_k = 59$ K for the emission component of $CH_3CN$ can be suitable. Even if $T_k$ of the emission component is varied, the radiation temperature of the absorption component is almost never changed since this is derived from relative intensities between $J = 4$–3 and 5–4.

The relative populations of individual rotational levels of $CH_3CN$ in this temperature condition are displayed in Figure 5. As a result, approximately 45% of this molecule belongs to the $J = K$ levels, at which rotations are restricted to being around the molecular axis. For example, in the case of the $K = 3$ stack, the $J = 3$ level has a much higher population compared with the $J = 4$ levels (Figure 5), whereas energies of both the levels come close to each other ($E_{J=3} = 48.4$ cm$^{-1}$ and $E_{J=4} = 50.8$ cm$^{-1}$). It is important to note that, if this molecule simply follows a Boltzmann distribution, the $J = K$ levels acquire only 2% of the total population at 88 K. These abundant populations of the $J = K$ levels create relatively strong absorption lines from the $J_K = 3_3$, $4_4$, and $5_5$ levels, according to the hot axis effect.

Since the kinetic temperatures of the moderate-density and low-density envelopes were suggested to be approximately 100 and >200 K, respectively (Hüttemeister et al. 1993, 1995), the kinetic temperature of $T_k = 88 \pm 29$ K ensures that the 64 km s$^{-1}$ component that includes $CH_3CN$ is in the envelope, not in a dense region of the Sgr B2(M) core with the kinetic temperature of $59 \pm 7$ K (Section 3.1). As a result, our observations suggest that a significant number of $CH_3CN$ molecules with $J = K$ level concentrations exists in the envelope of Sgr B2.

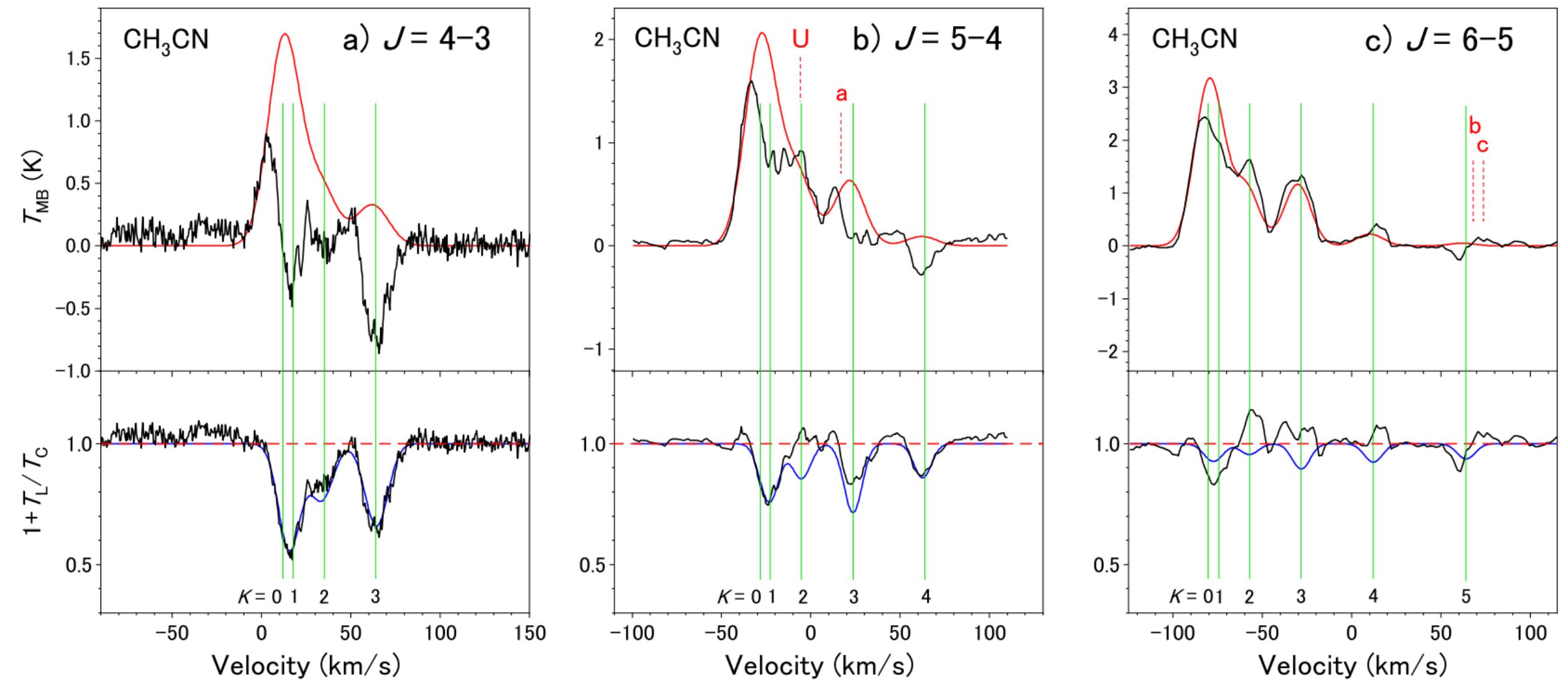

**Figure 4.** Absorption lines (black lines in lower panels) obtained by subtraction of emission lines of $CH_3CN$ toward Sgr B2(M). In the upper panels, the estimated emission profiles are depicted in red, and the observed profiles are in black. The velocities are determined based on the frequencies of the $J_K$ = $4_3$–$3_3$, $5_4$–$4_4$, and $6_5$–$5_5$ transitions, i.e., the transitions from the $J$ = $K$ levels, in (a), (b), and (c), respectively. The spectra of the $J$ = 5–4 and 6–5 transitions are from Belloche et al. (2013) by permission of ESO. The red bar marked by "a" in the middle trace shows a line of the $J$ = 5–4 transitions of $HC_3N$ $\nu_5$ = 1/$\nu_7$ = 3, and those marked by "b" and "c" in the right trace show lines of the $J_K$ = $5_0$–$4_0$ and $5_1$–$4_1$ transitions of $CH_3{}^{13}CN$ (CDMS). An unidentified line marked by "U" may exist at around 91,980 MHz, overlapping with $J_K$ = $5_2$–$4_2$. These lines cancel the absorption features of $CH_3CN$. In the lower panels, the blue lines are absorption profiles, reproduced using the derived temperatures and column density through least square fitting of the extracted absorption lines. Absorption profiles of the $J$ = 6–5 transitions are thought to be disturbed by some emission lines.

The extremely low $T_{ex}$ (3.0 K between the $J_K$ = $4_3$ and $3_3$ levels) of $CH_3CN$ suggests a contribution of the low-density envelope to the 64 km s$^{-1}$ component. However, further data must be collected to answer the questions about the actual distance between the 64 km s$^{-1}$ component and the Sgr B2(M) core. In addition, it is still uncertain whether the 64 km s$^{-1}$ component belongs to the moderate-density envelope or the low-density envelope. The present result does not contradict the suggestions that the 64 km s$^{-1}$ component is located close to the Sgr B2(M) core (Corby et al. 2015).

Pols et al. (2018) reported a 44 km s$^{-1}$ component toward Sgr B2(M) using the transitions $J$ = 12–11 to 14–13 of $CH_3CN$, attributing it to the spatially extended part of the envelope. Generally, higher rotational levels can be excited only in a high-temperature region. The rotational levels of this component reported by Pols et al. (2018) are higher than those in the present study, and their 44 km s$^{-1}$ component was not observed for the transitions $J$ = 4–3 to 6–5, as illustrated in Figure 4. Hence, we infer that the 44 km s$^{-1}$ component differs from the 64 km s$^{-1}$ component observed in this study. The temperature of the 44 km s$^{-1}$ component is higher than that of the 64 km s$^{-1}$ component. The observations of higher transitions by Pols et al. (2018) are thought to selectively detect the 44 km s$^{-1}$ component. Although the 44 km s$^{-1}$ component might generate relatively strong absorption lines from the $J$ = $K$ levels by the hot axis effect, Pols et al. (2018) did not present the lines in their figures, and the lines are out of the range of their analysis.

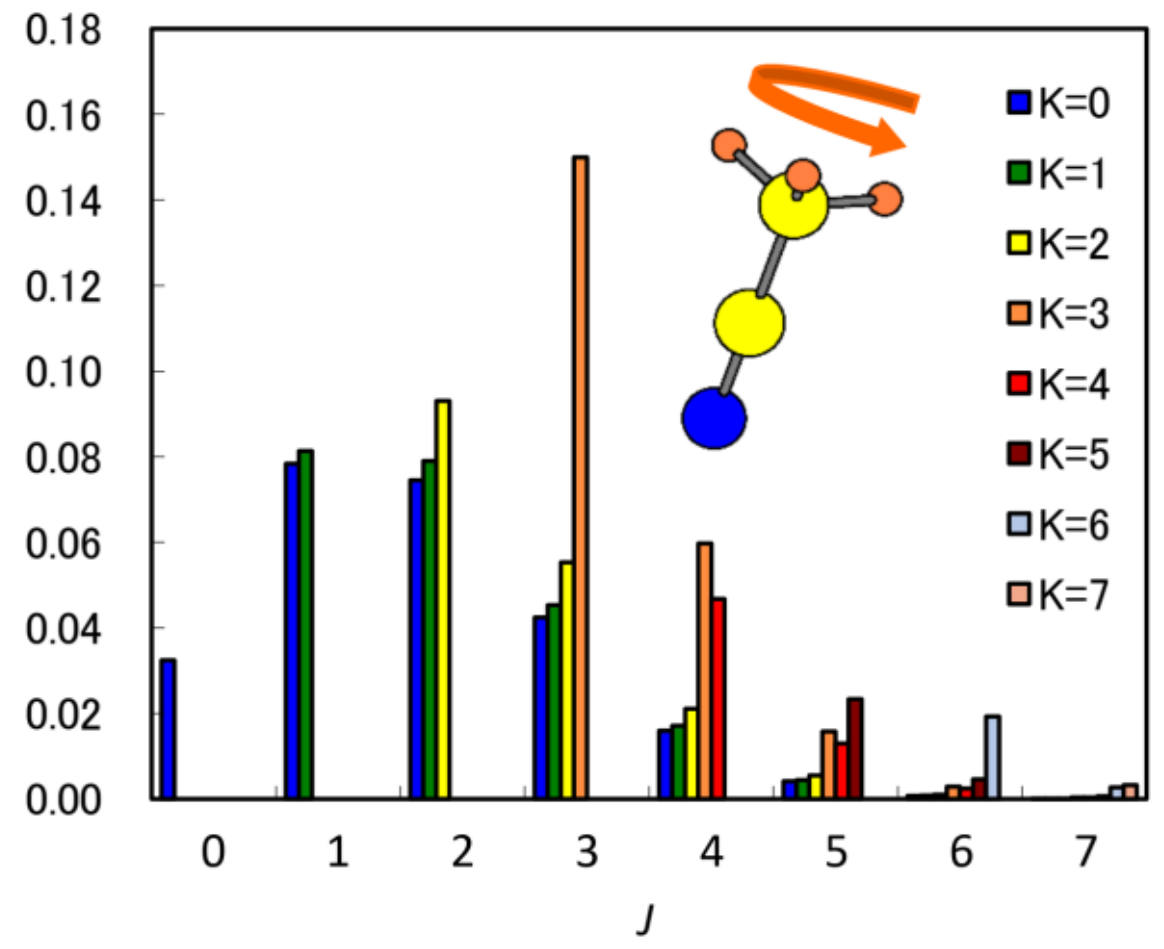

**Figure 5.** Relative populations of individual rotational levels by the *hot axis effect* (Appendix 1). $T_k$ = 88, $T_r$ = 2.8 K, and $C$ = 1 × 10$^{-6}$ s$^{-1}$ (Section 3.3). The summation of all populations is 1. Appendix 1 explains how to derive these populations. The illustration at the upper right shows a schematic image of rotation at the $J$ = $K$ levels of $CH_3CN$.

## 4. SUMMARY

To detect the $J = K$ level concentration of $CH_3CN$ in diffuse and translucent clouds, we observed absorption lines from $J = 4$–3 transitions at 73.6 GHz toward Sgr B2(M) using the Nobeyama Radio Observatory 45 m telescope. Both emission and absorption lines were found. By estimating the intensities of the emission lines from the Sgr B2(M) core, the absorption profiles from the envelope in the Sgr B2 region were extracted. A radiation temperature of 2.8 ± 0.5 K, a kinetic temperature of 88 ± 29 K, and column density of $(1.35 \pm 0.14) \times 10^{14}$ cm$^{-2}$ were derived for this molecule, and a low excitation temperature of 3.0 K was estimated between the $J_K = 4_3$ and $3_3$ levels. This suggests a hot axis effect, that is, the $J = K$ level concentrations caused by radiational cooling and excitation from collisions. The results confirm the existence of a significant number of $CH_3CN$ molecules concentrated in the $J = K$ levels in the relatively low-density region.

**Acknowledgments**

We thank the staff at Nobeyama Radio Observatory for help with the observations. Particularly, we are grateful to Dr. Hiroyuki Kaneko for his help with the continuum observations. M.A. thanks Grant-in-Aid for Scientific Research on Innovative Areas (Grant No. 25108002), Grant-in-Aid for Scientific Research (C) (Grant Nos. 15K05395 and 18K05045), and The Mitsubishi Foundation.

## APPENDICES

### A.1. ABSORPTION LINES OF $CH_3CN$

This section presents the theoretical basis of the hot axis effect. Under non-Boltzmann distribution in interstellar space, the number density $n(J)$ of each rotational level of a molecule depends on a radiational temperature, $T_r$, a kinetic temperature, $T_k$, optical depths of rotational transitions, and the rate of collision-induced rotational transitions, $C$. If no radiative sources exist around a cloud of relatively high density, an important factor is a set of optical depths rather than the radiational temperature of 2.7 K, the cosmic background temperature (Andersson et al. 1984). In this work, however, radiative sources exist around a cloud of relatively low density; therefore, a radiational temperature becomes an important variable to be determined by the following excitation calculations. We used a method based on Eq. (7) derived by Araki et al. (2014). This equation holds for $J > K_a$, i.e., $J$ = 4, 5, 6, … in the case of $K_a$ = 3, of a near-prolate molecule with $C_{2v}$ symmetry, and $n(0)$ in this equation means $n(K_a)$ to be exact. In the case of a symmetric-top molecule, the equation must be modified by replacing the rotational quantum number, $K_a$, with $K$, and the rotational constant, $\bar{B}\ (= (B + C^*)/2,\ C^*\text{: the rotational constant})$, with $B$ as follows:

$$n(J,K) = n(K,K) \prod_{m=K+1}^{J} \left[ \frac{\alpha B^3 \mu^2 \frac{m^3 S}{2m-1} \frac{1}{\exp(2hBm/kT_\mathrm{r}) - 1} + C\sqrt{\frac{2m+1}{2m-1}} \exp(-hBm/kT_\mathrm{k})}{\alpha B^3 \mu^2 \frac{m^3 S}{2m+1} \left(1 + \frac{1}{\exp(2hBm/kT_\mathrm{r}) - 1}\right) + C\sqrt{\frac{2m-1}{2m+1}} \exp(hBm/kT_\mathrm{k})} \right]$$

(A1)

for $J > K$, where $n(J, K)$ is the number density, $\alpha = 2^7 \pi^3/3\ \varepsilon_0\, h\, c^3$ is in SI units, $\varepsilon_0$ is the dielectric constant, $h$ is the Plank constant, $c$ is the speed of light, $k$ is the Boltzmann constant, $S$ is the transition strength, and $\mu$ is the permanent dipole moment in Cm (1 D = 3.33564 × $10^{-30}$ Cm). We note that an excitation temperature $T_{ex}$ of the $J$ rotation for each $K$ stack has the relationship $T_r < T_{ex} < T_k$, with $T_{ex}$ being closer to $T_k$ when $C$ is large and closer to $T_r$ when $C$ is small.

The population corresponding to the total column density, $N$, is distributed to all $K$ stacks based on a simple Boltzmann distribution at $T_k$. As a result, each $K$ stack obtains a column density, $N_K$, where

$$N = \sum_{K=0}^{\infty} N_K. \quad \text{(A2.1)}$$

The number densities, $n(J, K)$, for the respective $J$ levels in each $K$ stack are redistributed using Equation (A1). The column density, $N_{J,K}$, of each rotational level is

$$N_{J,K} = N_K\, n(J,K)/\sum_{j=K}^{\infty} n(j,K), \quad \text{(A2.2)}$$

where $n(K, K)$ in Equation (A1) is canceled by reduction.

Using the derived number densities, the intensity of an absorption line for a symmetric-top molecule in cgs units can be calculated by the conventional relationships between the column densities, $N_{J+1,K}$ and $N_{J,K}$, the optical depth, $\tau_{J+1,K\leftarrow J,K}$, of the $(J+1)_K \leftarrow J_K$ transition, and the Einstein B-coefficient:

$$\int \tau_{J+1,K\leftarrow J,K}\, dv = \frac{8\pi^3}{3h}\mu^2 S\left[\frac{N_{J,K}}{g_J} - \frac{N_{J+1,K}}{g_{J+1}}\right] \text{ in cm s}^{-1}\ (v\text{: velocity}), \quad \text{(A3)}$$

where $\mu$ is in Debye ($10^{-18}$ esu), $S = \frac{(J+1)^2 - K^2}{J+1}$, and $g_J = 2J+1$.

For $CH_3CN$ in diffuse and translucent clouds, the $J_K = 4_3$–$3_3$ transition produces the strongest absorption line, where $\mu$ = 3.9220 D (Gadhi et al. 1995) and $B$ = 9,198.9 MHz (Cazzoli & Puzzarini 2006). This is because: (1) the $J$ = 1–0, 2–1, and 3–2 transitions do not have sufficient population advantages at lower $J$ levels compared with upper ones; (2) the higher $J$ levels of $J \geq 4$ are disadvantaged compared with the $J$ = 3 level for obtaining populations; and (3) nuclear spin statistical weights of $K$ = 3n (n = 0, 1, 2, …) levels are twice those of $K \neq$ 3n. Therefore, we have selected the $J$ = 4–3 transition at 73.6 GHz to observe absorption lines of $CH_3CN$ in diffuse and translucent clouds. As a supplement, in transitions other than $J = K$, such as $J_K = 4_3$–$3_3$, the $K$ = 0 and 1 transitions in each $(J+1)_K - J_K$ transition can merge to create a single, strong absorption peak.

## A.2. ABSORPTION LINES OF $HC_3N$ TOWARD SGR B2(M)

For comparison in discussing the abundances of organic molecules, we focused on $HC_3N$ and $CH_3CN$. To obtain the column density of $HC_3N$ in the envelope of Sgr B2(M), we used the spectra of the $J$ = 4–3 line observed using the 100 m telescope at the Max Planck Institute for Radio Astronomy (MPIfR), reported by Hüttemeister et al. (1995). We also report in this work a new observation of the $J$ = 5–4 line using the 45 m telescope in Nobeyama. To detect this line at 45.5 GHz, a HEMT receiver H40 was used. The main beam efficiency was 0.69, and the beam size was 36.6″. A channel separation of 488.28 kHz with SAM45 was selected, giving an effective bandwidth of 1,600 MHz for each array. The corresponding velocity resolution is 6.43 km s$^{-1}$, sufficient to resolve the rotational profile of the line.

Rotational transitions of $HC_3N$ also exhibit absorption features at approximately 64 km s$^{-1}$, as displayed in Figure A1. Although absorption is dominant in the $J$ = 4–3 transition, the absorption and emission are mixed in the $J$ = 5–4 transition. To subtract emission lines from the core, the emission profiles of the $J$ = 4–3 and 5–4 transitions are estimated using common parameters of temperature, column density, full width at half maximum (FWHM), and velocity for $HC_3N$ in Table 86 of Belloche et al. (2013), as depicted by the red profiles in Figure A1. The profile of the lower velocity side of the $J$ = 5–4 lines, at which the absorption line has negligible overlap, is reproduced by the parameters, as illustrated in Figure A1 (b), as well as in the case of the $J_K = 4_0$–$3_0$ $CH_3CN$ line. Thus, the estimated emission profile of the $J$ = 4–3 transition of $HC_3N$ in Figure A1 (a) is suitable.

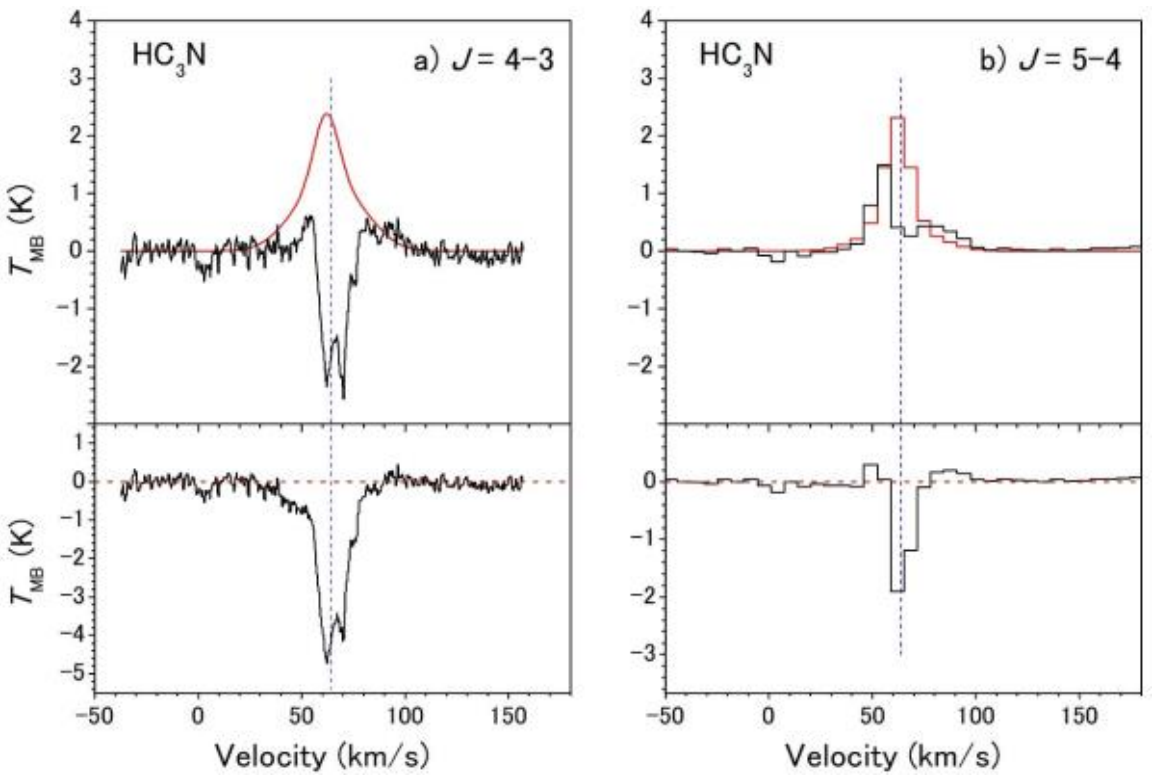

**Figure A1.** Absorption lines (lower panels) of $HC_3N$. The $J$ = 4–3 line was observed using the 100 m telescope of the MPIfR by Hüttemeister et al. (1995). This spectrum is reprinted from their paper with permission of ESO. The $J$ = 5–4 line was obtained using the Nobeyama 45 m telescope in this work. The black lines in the upper panels are the observed spectra. The red lines in the upper panels are the expected emission profiles using the parameters of $HC_3N$ in Table 86 of Belloche et al. (2013). The lower panels depict the absorption profiles after subtracting the emission profiles. The blue dotted lines are at 64 km s$^{-1}$.

The absorption profiles of $HC_3N$ in the lower panels of Figure A1 are obtained by removing the emission profiles. The background continuum intensities, $T_C$, are 21 K at 36.4 GHz, corresponding to the $J$ = 4–3 transition, and 14 K at 43.4 GHz, observed using the 100 m telescope (Hüttemeister et al. 1995). The intensity of $T_C$ = 14 K can be used as the upper limit of $T_C$ at 45.5 GHz for the $J$ = 5–4 transition with t he 45 m telescope, since the beam size of the 45 m telescope is larger than that of the 100 m instrument. The $J$ = 5–4 absorption line gives the lower limit (>3 × $10^{13}$ cm$^{-2}$) of the column density of $HC_3N$ according to the local thermodynamic equilibrium equation for absorption lines of a linear molecule described by Greaves et al. (1992). Using the absorption intensity of the $J$ = 4–3 transition, a column density of 4 × $10^{13}$ cm$^{-2}$ is derived based on the assumption of $T_{ex}$ = 2.9 K, which is $T_r$ for $CH_3CN$. This column density, which varies by 5% per 1 K change of $T_C$ and by 40% per 1 K increase of $T_{ex}$, is consistent with the lower limit obtained from the $J$ = 5–4 transition.

### A.3. ABUNDANCES OF ORGANIC MOLECULES

To understand the history of organic molecules in space, we compare the abundances of organic molecules, including COMs in different several clouds. In a previous work, Liszt et al. (2018) compared the abundances of five dominant species, $H^{13}CO^+$, *c*-$C_3H_2$, $CH_3OH$, $CH_3CN$, and $HC_3N$, in one dense cloud and several diffuse and translucent clouds; they suggested that the abundances of the organic molecules in diffuse and translucent clouds may be mutually different. We extend this comparison to column density of $H^{13}CO^+$ (Liszt et al. 2018). The densities are listed in Table A1 and displayed in Figure A2. Despite the difference in absolute densities between the envelope and the Sgr B2(M) core, the relative abundances of COMs are similar. Thus, the abundances of COMs in a relatively low-density region in Sgr B2(M) are thought to be comparable to those in the dense core. Price et al. (2003), Garrod et al. (2005, 2006), and Thiel et al. (2017) have suggested that interstellar matter is exchanged among dense, translucent, and diffuse clouds. The similarity of COM abundances between the envelope and the Sgr B2(M) core can be consistent with this model of interstellar-matter exchange. Price et al. (2003) suggested that diffuse clouds produced from diffuse atomic clouds and those produced from dense clouds demonstrate clear

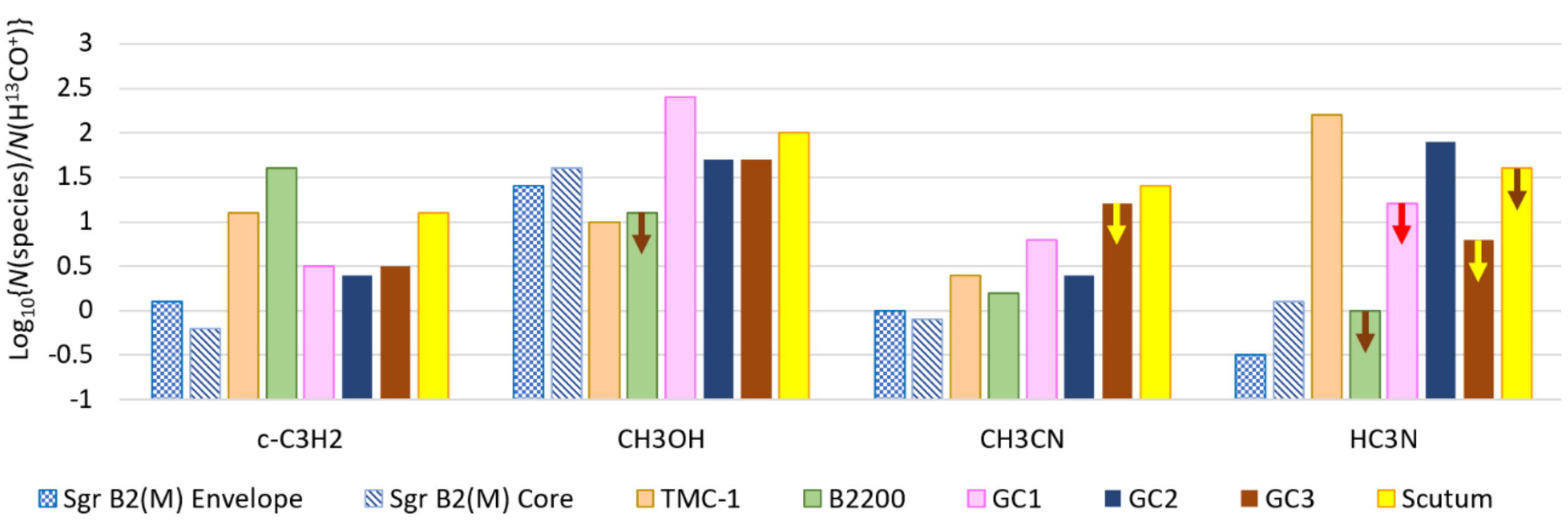

**Figure A2.** Comparison of abundances among dense and translucent clouds. All values come from Table A1. Arrows show upper limits.

the envelope and the core of Sgr B2 (M).

To compare the abundances of organic molecules based on column-density ratios, the column densities of the organic molecules are normalized by those of $H^{13}CO^+$ because the column density of $H_2$ can be related to the differences in chemical composition. This proposal may be supported by the variety of COM abundances among the envelope and diffuse and translucent clouds presented in Table A1 and Figure A2.

**Table 2**
Beam Sizes and Continuum Intensities $T_C$

| | Molecule (continuum) | Transition $J$ | Frequency (GHz) | Beam size (″) | $\eta_B$ | Telescope | $T_C$ (K) |
|---|---|---|---|---|---|---|---|
| B0212+735 | continuum | | 80 | 18.7 | 0.58 | NRO 45 m | 1.07(17) [a] |
| Sgr B2(M) | $CH_3CN$ | 6–5 | 110.3 | 22.3 | | IRAM 30 m | 2.7(7) [b,c] |
| | $CH_3CN$ | 5–4 | 92.0 | 26.8 | | IRAM 30 m | 2.7(7) [b,c] |
| | $CH_3CCH$ | 5–4 | 85.4 | 19.1 | 0.55 | NRO 45 m | 2.7(7) [b] |
| | continuum | | 80 | 18.7 | 0.58 | NRO 45 m | 2.7(7) [d] |
| | $CH_3CN$ | 4–3 | 73.6 | 20.4 | 0.55 | NRO 45 m | 2.7(7) [b] |
| | $HC_3N$ | 5–4 | 45.5 | 36.6 | 0.69 | NRO 45 m | <14 |
| | SiO | 1–0 | 43.4 | 23 | | MPIfR 100 m | 14 [e] |
| | $HC_3N$ | 4–3 | 36.4 | 27 | | MPIfR 100 m | 21 [e] |
| W49N | continuum | | 80 | 18.7 | 0.58 | NRO 45 m | 2.80(9) [a] |
| W51 | continuum | | 80 | 18.7 | 0.58 | NRO 45 m | 2.78(9) [a] |

**Note:** The numbers in the parentheses are the errors in units of the last significant digit. $T_C$ does not include the cosmic background radiation (e.g., Greaves et al. 1992).
[a] The same value is assumed as an intensity at 73.6 GHz.
[b] Estimated from the observed flux density at 80.0 GHz assuming a constant value for the flux density in this frequency region.
[c] This value is used to evaluate $T_L/T_C$ from the reported spectrum by Belloche et al. (2013).
[d] Observed in this work.
[e] Hüttemeister et al. (1995).

**Table 3**

Emission Lines of the $J_K = 5_K$–$4_K$ Rotational Transitions of $CH_3CCH$

| Source | Transition $K$ | Frequency [a,b] (GHz) | $T_{MB}$ [c] (K) | $\Delta v$ [c] (km s$^{-1}$) | $W$ [d] (K km s$^{-1}$) | $V_{LSR}$ (km s$^{-1}$) | rms (mK) |
|---|---|---|---|---|---|---|---|
| B0212+735 | 0 | 85457.300 | <0.09 | | | | 31 |
| | 1 | 85455.667 | <0.09 | | | | |
| | 2 | 85450.766 | <0.09 | | | | |
| | 3 | 85442.601 | <0.09 | | | | |
| | 4 | 85431.175 | <0.09 | | | | |
| Orion IRc2 | 0 | | 1.11(8) | 3.5(3) | 4.1(5) | 8.7 | 36 |
| | 1 | | 0.88(8) | 3.7(5) | 3.5(5) | 8.7 | |
| | 2 | | 0.53(7) | 3.7 | 2.1(3) | 8.7 | |
| | 3 | | 0.42(7) | 3.7 | 1.7(3) | 8.6 | |
| SgrB2 (M) | 0 | | 0.97(46) | 13.2 | 13.7(65) [e] | 61.6 | 25 |
| | 1 | | 0.70(44) | 13.2 | 9.9(62) [e] | 62.7 | |
| | 2 | | 0.52(6) | 13.2 | 7.3(9) | 62.1 | |
| | 3 | | 0.43(5) | 13.2 | 6.1(8) | 61.4 | |
| W49N | 0 | | 0.251(4) | 6.6 | 1.77(3) | 3.6 | 10 |
| | 1 | | 0.167(9) | 6.6 | 1.17(6) | 3.6 | |
| | 2 | | 0.132(4) | 6.6 | 0.93(3) | 3.6 | |
| | 3 | | 0.120(4) | 5.8 [d] | 0.74(3) | 3.6 | |
| | 4 | | 0.022(4) | 5.8 [d] | 0.13(3) | 3.6 | |
| | 0 | | 0.350(9) | 6.6 | 2.46(6) | 11.5 | 10 |
| | 1 | | 0.283(4) | 6.6 | 1.99(3) | 11.5 | |
| | 2 | | 0.158(4) | 6.6 | 1.11(3) | 11.5 | |
| | 3 | | 0.124(4) | 5.8 [d] | 0.76(3) | 11.5 | |
| | 4 | | 0.014(4) | 5.8 [d] | 0.09(3) | 11.5 | |
| W51 | 0 | | 0.92(5) | 6.0(3) | 5.8(4) | 56.3 | 29 |
| | 1 | | 0.88(4) | 6.7(4) | 6.3(4) | 56.5 | |
| | 2 | | 0.418(14) | 7.5(3) | 3.34(17) | 56.5 | |
| | 3 | | 0.328(14) | 7.4(3) | 2.57(17) | 56.4 | |

**Note:** The numbers in the parentheses are the one-sigma errors in units of the last significant digit.

[a] Rest frequency, which is common in all clouds.

[b] The Cologne Database for Molecular Spectroscopy (Müller et al. 2001, 2005).

[c] Main beam temperature $T_{MB}$ and FWHM are obtained by Gaussian fit. A value having no error is fixed in the fit.

[d] $W = \int T_{MB} dv$.

[e] Large uncertainties are due to the blending of the two transitions. Sums of the integrated intensities of both the lines have small uncertainties.

[d] Narrower widths are good for fitting.

**Table 4**
Temperatures and Column Densities of the Observed Species

| Species | | Source | Kinetic Temperature $T_k$ (K) | Temperature [a] $T_{ex}$ or $T_r$ (K) | Column Density $N$(cm$^{-2}$) | |
|---|---|---|---|---|---|---|
| $CH_3CCH$ | emission | Orion IRc2 | 50.5(5) | 24 [b] | 8.10(6) | × $10^{14}$ |
| | | Sgr B2(M) | 59(7) | 30 [b] | 2.36(16) | × $10^{15}$ |
| | | W49N 3.5 km s$^{-1}$ | 68(7) | 31 [c] | 1.70(16) | × $10^{14}$ |
| | | 11.5 km s$^{-1}$ | 45.1(10) | 31 [c] | 2.23(7) | × $10^{14}$ |
| | | W51 | 48(6) | 20 [d] | 1.19(10) | × $10^{15}$ |
| $CH_3CN$ | emission | Orion IRc2 [e] | 270(170) | 32.4 [f] | 1.14(11) | × $10^{15}$ |
| | | Sgr B2(M) | 59 [g] | 16 [h] | 1.7 | × $10^{14}$ |
| | | Sgr B2(M) edge [i] | 72(18) | 16 [h] | 7.1(15) | × $10^{13}$ |
| | | W49N | 68(14) | 31 [c] | 3.3(4) | × $10^{13}$ |
| | | W51 | 122(35) | 12 [j] | 1.35(11) | × $10^{14}$ |
| | absorption | toward B0212+735 [k] | 100 [l] | 2.7 [l] | <4 | × $10^{13}$ |
| | | toward Sgr B2(M) [m] | 88(29) | 2.8(5) [n] | 1.35(14) | × $10^{14}$ |
| | | toward W49N [k] | 100 [l] | 2.7 [l] | <7 | × $10^{12}$ |
| | | toward W51 [k] | 100 [l] | 2.7 [l] | <2 | × $10^{13}$ |

**Note:** The numbers in parentheses are the one-sigma errors in units of the last significant digit.
[a] An excitation temperature, $T_{ex}$, in the case of emission and radiation temperature, $T_r$, in the case of absorption.
[b] Fixed at the temperature reported by Churchwell and Hollis (1983).
[c] Fixed at the temperature of $CH_3CCH$ (Nagy et al. 2015).
[d] Fixed at the temperature of $O^{13}CS$ (Kalenskii & Johansson 2010).
[e] The $K$ = 0 line was not in the fitting. Single component analysis is not appropriate.
[f] Fixed at the temperature of $HC_3N$ (Turner 1991).
[g] Fixed at the temperature of $CH_3CCH$ in Sgr B2(M) in this work.
[h] Fixed at the temperature of $CH_3CN$ (Cummins et al. 1983).
[i] Side position of Sgr B2(M). The column density is the sum of $K$-ladders, $K$ = 0–5.
[j] Fixed at the temperature of $HC^{13}CCN$ (Kalenskii & Johansson 2010).
[k] A rate for collision-induced rotational transitions of $10^{-7}$ s$^{-1}$ is used, based on the assumption of a diffuse cloud.
[l] Fixed.
[m] 64 km s$^{-1}$ component. A rate for collision-induced rotational transitions of $10^{-6}$ s$^{-1}$ is assumed (see Section 3.3).
[n] $T_{ex}$ between the $J_K = 4_3$ and $3_3$ levels, estimated to be 3.0 K.

**Table 5**
Emission Lines of $CH_3CN$

| Transition $J$ | Source | Transition $K$ | Frequency [a,b] (GHz) | $T_{MB}$ (K) | $\Delta v$ [c] (km s$^{-1}$) | $W$ [d] (K km s$^{-1}$) | $V_{LSR}$ (km s$^{-1}$) | rms (mK) |
|---|---|---|---|---|---|---|---|---|
| 4–3 | B0212+735 | 0 | 73590.219 | <0.17 | | | | 56 |
| | | 1 | 73588.800 | <0.17 | | | | |
| | | 2 | 73584.543 | <0.17 | | | | |
| | | 3 | 73577.452 | <0.17 | | | | |
| | Orion IRc2 | 0 | | 5.77(14) [e] | 8.6 | 53.0(12) | 6.1 | 78 |
| | | 1 | | 5.93(13) | 8.6 | 54.5(12) | 7.5 | |
| | | 2 | | 4.52(9) | 9.9(2) | 47.6(14) | 6.5 | |
| | | 3 | | 3.53(8) | 12.5(3) | 46.8(16) | 6.4 | |
| | Sgr B2(M) [f] | 0 | | 0.950 | 19.8 | 20.1 | 62 | 62 |
| | | 1 | | 0.811 | 19.8 | 17.1 | 62 | |
| | | 2 | | 0.461 | 21.1 [g] | 10.3 | 62 | |
| | | 3 | 73577.948 [h] | 0.228 | 21.1 [g] | 5.1 | 62 | |
| | | 3 | 73576.501 [i] | 0.117 | 19.8 | 2.5 | 62 | |
| | W49N | 0 | | 0.221(10) | 9.6 | 2.26(11) | 6.7 | 24 |
| | | 1 | | 0.250(10) | 9.6 | 2.56(11) | 9.3 | |
| | | 2 | | 0.121(6) | 11.5 | 1.49(7) | 8.6 | |
| | | 3 | | 0.060(6) | 15(2) | 1.0(2) | 7.3 | |
| | W51 | 0 | | 1.37(7) | 9.6 | 14.0(7) | 56.5 | 66 |
| | | 1 | | 1.02(7) | 9.6 | 10.4(7) | 56.9 | |
| | | 2 | | 0.94(2) | 9.5(3) | 9.5(3) | 56.3 | |
| | | 3 | | 0.67(2) | 12.8(4) | 9.2(4) | 56.7 | |
| 6–5 | Sgr B2(M) Side Position | 0 | 110383.500 | 0.59(2) | 19 | 12.0(4) | 62 [h] | 13 |
| | | 1 | 110381.372 | 0.26(2) | 19 | 5.4(4) | 62 [h] | |
| | | 2 | 110374.989 | 0.361(10) | 19 | 7.4(2) | 62 [h] | |
| | | 3 | 110364.354 | 0.282(9) | 21(1) | 6.3(3) | 62 [h] | |
| | | 4 | 110349.471 | 0.077(10) | 20(3) | 1.7(3) | 62 [h] | |
| | | 5 | 110330.345 | 0.037(10) | 18(6) | 0.7(3) | 62 [h] | |

**Note:** The numbers in the parentheses are the one-sigma errors in units of the last significant digit.
[a] Rest frequency, which is common in all clouds. Average of hyperfine components.
[b] The Cologne Database for Molecular Spectroscopy (Müller et al. 2001, 2005) and the paper of Boucher et al. (1980).
[c] FWHM obtained by Gaussian fit. A value having no error is fixed in the fit.
[d] $W = \int T_{MB} dv$.
[e] The $K = 0$ line was not in the fitting. Single component analysis is not appropriate because of the broad width.
[f] Estimated intensities. See Section 3.2.
[g] A large width is assumed because of the hyperfine components.
[h] $F = 5$–4 and 3–2.
[i] $F = 4$–3.

**Table 6**
Absorption Lines of $CH_3CN$

| Line of sight | $J$ | $K$ | $V_{LSR}$ (km s$^{-1}$) | $\|T_L/T_C\|$ [a] | $\Delta v$ [b] (km s$^{-1}$) | rms (mK) |
|---|---|---|---|---|---|---|
| B0212+735 | 4–3 | 3 | - | <0.155 | - | 56 |
| Sgr B2(M) | 4–3 | 0 & 1 [c] | - | 0.457(8) | 14.4(5) | 62 |
| | | 2 | 62.5(6) | 0.176(7) | 15.3(13) | |
| | | 3 | 64.34(15) | 0.363(7) | 15.1(4) | |
| | 5–4 | 0 & 1 [c] | - | 0.245(9) | 12.4(5) | 18 |
| | | 2 [d] | - | - | - | |
| | | 3 | 65.2(3) | 0.169(8) | 12.0(7) | |
| | | 4 | 62.6(3) | 0.136(9) | 11.5(8) | |
| | 6–5 | 0 and 1 [c] | - | 0.168(15) | 11.7(11) | 22 |
| W49N | 4–3 | 3 | - | <0.026 | - | 24 |
| W51 | 4–3 | 3 | - | <0.072 | - | 66 |

**Note:** The upper limits of $|T_L/T_C|$ for Orion IRc2 and the side position of Sgr B2(M) cannot be estimated because suitable data of continuum intensities are not available. The numbers in the parentheses are the one-sigma errors in units of the last significant digit.
[a] Three times the rms was assumed as an upper limit of detection.
[b] FWHM obtained by the Gaussian fit.
[c] Blend.
[d] A U-line may exist at around 91,980 MHz.

**Table 7**
Rates for Collision-Induced Rotational Transitions in s$^{-1}$ between $CH_3CN$ and $H_2$

| $H_2$ density | Kinetic Temperature | | |
|---|---|---|---|
| cm$^{-3}$ | 300 K | 100 K | 10 K |
| $10^2$ | $1.5 \times 10^{-7}$ | $1 \times 10^{-7}$ | $3 \times 10^{-8}$ |
| $10^3$ | $1.5 \times 10^{-6}$ | $1 \times 10^{-6}$ | $3 \times 10^{-7}$ |
| $10^5$ | $1.5 \times 10^{-4}$ | $1 \times 10^{-4}$ | $3 \times 10^{-5}$ |
| $10^7$ | $1.5 \times 10^{-2}$ | $1 \times 10^{-2}$ | $3 \times 10^{-3}$ |

**Note:** Rates for collision-induced rotational transitions between $CH_3CN$ and $H_2$ at room temperature at the pressure of 3–5 mTorr were experimentally obtained in errors of 6% by Mäder et al. (1979). These rates are suitable for deriving rates for the low-$J$ rotational transitions observed in this work, because these rates were measured by using the $J$ = 1–0 and 2–1 rotational transitions. To estimate rates in space conditions, a hard-sphere collisional model is used. The $H_2$ densities of $10^2$, $10^3$, $10^5$, and $10^7$ cm$^{-3}$ may correspond to a typical diffuse cloud, the low-density envelope, the moderate-density envelope, and the Sgr B2(M) core, respectively. Green (1986) derived the theoretical collision rates between $CH_3CN$ and $H_2$ as listed in Leiden Atomic and Molecular Database (Schöier et al. 2005). These theoretical rates were reported to be in accord with the values of the room temperature measurements (Mäder et al. 1979) and agree with our rates within one order of magnitude at 100 K. The assumed rate in the present analysis is discussed in Section 3.3.

**Table A1**

Comparison of Abundances in Dense, Translucent, and Diffuse Clouds

| Species | Sgr B2(M) Envelope [c] | Sgr B2(M) Core [d] | TMC-1 CP | toward B2200 [a] | GC1 [b] | GC2 [b] | GC3 [b] | Scutum [b] |
|---|---|---|---|---|---|---|---|---|
| | Column density ($cm^{-2}$) | | | | | | | |
| $H^{13}CO^+$ | $1.4 \times 10^{14}$ [e] | $3.72 \times 10^{14}$ [f] | $1.45 \times 10^{12}$ [g] | $4.2 \times 10^{10}$ | $1.5 \times 10^{12}$ | $8.0 \times 10^{12}$ | $3.8 \times 10^{12}$ | $6.2 \times 10^{11}$ |
| c-$C_3H_2$ | $1.9 \times 10^{14}$ [e] | $2.60 \times 10^{14}$ [f] | $1.9 \times 10^{13}$ [h] | $1.5 \times 10^{12}$ | $5.0 \times 10^{12}$ | $1.9 \times 10^{13}$ | $1.2 \times 10^{13}$ | $8.1 \times 10^{12}$ |
| $CH_3OH$ | $3.83 \times 10^{15}$ [f] | $1.63 \times 10^{16}$ [f] | $1.4 \times 10^{13}$ [h] | $<5 \times 10^{11}$ | $3.8 \times 10^{14}$ | $3.8 \times 10^{14}$ | $2.1 \times 10^{14}$ | $6.2 \times 10^{13}$ |
| $CH_3CN$ | $1.4 \times 10^{14}$ [i] | $2.69 \times 10^{14}$ [f] | $4.1 \times 10^{12}$ [h] | $7 \times 10^{10}$ | $1.0 \times 10^{13}$ | $2.0 \times 10^{13}$ | $<6.0 \times 10^{13}$ | $1.4 \times 10^{13}$ |
| | | $1.7 \times 10^{14}$ [i] | | | | | | |
| $HC_3N$ | $4 \times 10^{13}$ [j] | $4.50 \times 10^{14}$ [f] | $2.3 \times 10^{14}$ [h] | $<4 \times 10^{10}$ | $<2.5 \times 10^{13}$ | $6.0 \times 10^{14}$ | $<2.5 \times 10^{13}$ | $<2.5 \times 10^{13}$ |
| | $Log_{10}\{N(species)/N(H^{13}CO^+)\}$ | | | | | | | |
| $H^{13}CO^+$ | 0.0 | 0.0 | 0.0 | 0.0 | 0.0 | 0.0 | 0.0 | 0.0 |
| c-$C_3H_2$ | 0.1 | −0.2 | 1.1 | 1.6 | 0.5 | 0.4 | 0.5 | 1.1 |
| $CH_3OH$ | 1.4 | 1.6 | 1.0 | <1.1 | 2.4 | 1.7 | 1.7 | 2.0 |
| $CH_3CN$ | 0.0 | −0.1 | 0.4 | 0.2 | 0.8 | 0.4 | <1.2 | 1.4 |
| | | −0.3 | | | | | | |
| $HC_3N$ | −0.5 [j] | 0.1 | 2.2 | <0.0 | <1.2 | 1.9 | <0.8 | <1.6 |

**Notes.**

[a] Liszt et al. (2018).
[b] Galactic Center translucent clouds reported by Thiel et al. (2017). GC1, GC2, and GC3 are distinguished by their velocities of approximately 9, 3, and −1 km $s^{-1}$, respectively.
[c] 64 ± 1 km $s^{-1}$. The moderate density envelope observed by absorption.
[d] 62 ± 1 km $s^{-1}$. Core observed by emission.
[e] Menten et al. (2011).
[f] Belloche et al. (2013).
[g] Pratap et al. (1997).
[h] Gratier et al. (2016).
[i] This work.
[j] $T_{ex}$ = 2.9 K.